# Representation reduction and solution space contraction in quasi-exactly solvable systems

A. D. Alhaidari

*Shura Council, Riyadh 11212, Saudi Arabia*
E-mail: haidari@mailaps.org

In quasi-exactly solvable problems partial analytic solution (energy spectrum and associated wavefunctions) are obtained if some potential parameters are assigned specific values. We introduce a new class in which exact solutions are obtained at a given energy for a special set of values of the potential parameters. To obtain a larger solution space one varies the energy over a discrete set (the spectrum). A unified treatment that includes the standard as well as the new class of quasi-exactly solvable problems is presented and few examples (some of which are new) are given. The solution space is spanned by discrete square integrable basis functions in which the matrix representation of the Hamiltonian is tridiagonal. Imposing quasi-exact solvability constraints result in a complete reduction of the representation into the direct sum of a finite and infinite component. The finite is real and exactly solvable, whereas the infinite is complex and associated with zero norm states. Consequently, the whole physical space contracts to a finite dimensional subspace with normalizable states.

## I. INTRODUCTION

One may choose not to disagree with the view that *exactly* solvable systems are, by some (debatable) definitions, trivial. Nonetheless, an advantage of obtaining exact solutions of the wave equation is that the analysis of such solutions makes the conceptual understanding of physics straightforward and sometimes intuitive. Moreover, these solutions are valuable means for checking and improving models and numerical methods introduced for solving complicated physical systems. In fact, in some limiting cases or for some special circumstances they may constitute analytic solutions of realistic problems or approximations thereof. In the exactly solvable class of problems, full analytic solutions (energy spectrum and associated state functions) could be obtained for a continuous range of values of the physical parameters of the system. However, elements of this class must be endowed with a high degree of symmetry and are, in fact, very limited in number. Most of the known exactly solvable problems fall within distinct classes of, what is referred to as, "shape invariant potentials" [1]. Each class carries a representation of a given symmetry group. Elements of the solutions in each class could be transformed into one another by the action of the operators of the associated symmetry algebra. Supersymmetric quantum mechanics, potential algebras, point canonical transformations, and path integration are four methods among many which are used in the search for exact solutions of the wave equation. In nonrelativistic quantum mechanics, these developments were carried out over the years by many researchers where several of these solutions are accounted for and tabulated (see [1] and references cited therein). This class includes dynamical systems with potentials like the Coulomb, Oscillator, Morse, Pöschl-Teller, Hulthén, Scarf, etc.

A larger class of problems is exactly solvable only if the physical parameters of the system are assigned specific values. Two such classes exist. In the *conditionally-exactly* solvable class all energy eigenvalues and corresponding wavefunctions are obtained exactly under the given constraint on the physical parameters [2]. However, in the *quasi-exactly* solvable class of problems only partial solution is possible [3]. That is, only part of the energy spectrum and associated wavefunctions are obtained exactly under the given parameters constraint. In this work we introduce a new type of solutions to the latter class where the relationship between the potential parameters and the energy spectrum is interchanged. In other words, partial solution is obtained at a specific energy for a large set of special values of the potential parameters (referred to, hereafter, as the "parameter spectrum"). Subsequently, we propose a unified treatment that covers both classes of quasi-exactly solvable problems in which either the potential parameters are fixed and solutions are obtained for a set of values of the energy (energy spectrum) or the energy is fixed and solutions are obtained for a set of values of the potential parameters (the "parameter spectrum"). We show that each problem in either class is associated with a set of polynomials, in the energy or potential parameters, that satisfy a three-term recursion relation. These polynomials are the expansion coefficients of the wavefunction in a given square integrable basis. However, not all energy polynomials form orthogonal sets with respect to a positive energy measure. We also show that solvability requirements of these problems result in a complete reduction of the representation into the direct sum of a finite and infinite component. The finite one is real and solvable, whereas the infinite is complex and has zero norm states. Several examples will be given to illustrate the utility of the approach. Some of these examples involve potentials that were not treated before.

Two types of recursion relations for the energy polynomials are obtained. We associate with each one a class of solutions. One will be referred to as the "diagonal representation" class and the other as the "off-diagonal representation". In the three-term recursion relation associated with the former class the energy variable appears, as usual, in the central term of the recursion, whereas in the latter it appears in either one of the two end terms. The usual class of quasi-exactly solvable problems belongs to either of the two representations whereas the new one introduced here belongs only to the off-diagonal representation. Additionally, the latter class is further divided into two subclasses corresponding to either of the two locations of the energy variable in the recursion relation. For a given choice of $L^2$ basis, it might be possible that some of these classes or subclasses are empty. In the diagonal representation, two quasi-exact solvability conditions are obtained. Each one gives a different constraint on the parameters of the problem for obtaining exact partial solutions. In the off-diagonal representation, one of the two solvability constraints limits the quasi-exact solution to only one value of the energy but leaves more freedom for the problem parameters to vary within the "parameter spectrum". To obtain a larger solution space in the new class one could vary the energy over a special discrete set.

The general formulation for the proposed unified treatment of quasi-exact solvability will be presented in the following section. Starting with a suitable basis, general conditions are imposed such that the wave equation results in a three-term recursion relation for the expansion coefficients of the wavefunction. Consequently, two different representations emerge. In the diagonal representation, the energy appears linearly in the central term of the recursion. However, in the off-diagonal representation it appears in one of the two end terms. Thereafter, and in an essential and critical step of the approach, a transformation of the basis is performed such that the matrix representation of the Hamiltonian (or its equivalent) becomes tridiagonal and symmetric. Subsequently, a condition is imposed on the parameters of the problem such that the matrix representation becomes completely reducible into a finite and infinite part. General reality and solvability constraints on the finite representation are found. Detailed treatment of the diagonal representation will be carried out in Sec. III where we also illustrate explicitly the contraction of the solution space. Two sets of examples are given including new problems that have never been treated before (e.g., a partner to the Bender-Dunne potential [4]). The *S*-wave Morse potential plus an exponentially rising term is also one of the problems treated in the same section. One of the two subclasses in the off-diagonal representation will be investigated in Sec. IV where the solvability constraint results in a solution at only one value of the energy but for several special values of the potential parameters. This finding introduces the concept of the "parameter spectrum". The energy polynomials (expansion coefficients of the wavefunction) do not form an orthogonal set with respect to a *positive* measure. One of the examples given includes a problem with the oscillator potential plus the inverse-quartic potential. Another includes a sum of the Coulomb and oscillator potentials. The other subclass in the off-diagonal representation will be treated in Sec. V. The potential example given there is a generalization of the Morse potential which includes not just the exponential term and its square but its half-power as well. In the Appendix extra examples are listed without giving the details.

## II. GENERAL FORMULATION

The time-independent Schrödinger equation for a scalar particle of mass *m* in the field of a potential $V(x)$ reads as follows

$$\left[-\frac{1}{2}\frac{d^2}{dx^2}+V(x)-E\right]\psi_E(x)=0\,, \tag{2.1}$$

where $x\in[x_-,x_+]\subseteq\Re$ and we have used the atomic units $\hbar=m=1$. This is the steady state quantum mechanical equation in one dimension. It could also be considered as the radial *S*-wave Schrödinger equation for a particle in spherically symmetric potential where $x$ stands for the radial coordinate and where $x_-=0$ and $x_+=\infty$. Moreover, if $V(x)$ contains a term that could be written as $\ell(\ell+1)/2x^2$ with $\ell=0,1,2,..$, then Eq. (2.1) could also be interpreted as describing the dynamics in a spherically symmetric potential with orbital angular momentum $\ell$. Now, for static (steady-state) probability density, the full wavefunction belongs to the space of square integrable functions. In other words, we can always write it as a sum in an $L^2$ basis functions, $\{\phi_n(x)\}_{n=0}^{\infty}$, that are compatible with the domain of the Hamiltonian $H$, where $H=-\frac{1}{2}\frac{d^2}{dx^2}+V(x)$. That is, we can write $\psi_E(x)$ = $\sum_{n=0}^{\infty}p_n\,\phi_n(x)$, where the expansion coefficients $p_n$ are energy-dependent. With a proper choice of weight function $\omega(x)$ we can almost always write the basis function as $\phi_n(x)=\omega(y)y^{\mu n}$ for a suitable coordinate transformation $y(\lambda x)$, where $\mu$ and $\lambda$ are real parameters and $\lambda$ positive with inverse length dimension. Square integrability requires that $\int_{x_-}^{x_+}\phi_n^2(x)\,dx=\frac{1}{\lambda}\int_{y_-}^{y_+}\frac{1}{y'}\omega^2y^{2\mu n}dy$ be finite for all $n$, where the prime on $y$ means the derivative with respect to *x*. We define $h=2\lambda^{-2}H$, $v=2\lambda^{-2}V$, and $\varepsilon=2\lambda^{-2}E$. For simplicity and economy of notation we choose $\lambda$ to be the unit of length. Thus, the action of the wave operator $H-E$ on the basis could be written as

$$\begin{aligned}\left(-h+\varepsilon\right)\phi_n &= \left(y'^2\tfrac{d^2}{dy^2}+y''\tfrac{d}{dy}-v+\varepsilon\right)\phi_n \\ &= \omega\left[y'^2\tfrac{d^2}{dy^2}+\left(y''+2y'^2\tfrac{\omega'}{\omega}\right)\tfrac{d}{dy}+y'^2\tfrac{\omega''}{\omega}+y''\tfrac{\omega'}{\omega}-v+\varepsilon\right]y^{\mu n}\end{aligned} \tag{2.2}$$

where the prime on $\omega$ stands for the derivative with respect to its argument, *y*. Therefore, we can rewrite the wave equation (2.1) as follows

$$\omega(y)\sum_{n=0}^{\infty}y^{\mu n}\left[\mu n(\mu n-1)A(y)+\mu nB(y)+C(y)-v(y)+\varepsilon\right]p_n(\varepsilon)=0\,, \tag{2.3}$$

where $A(y)=(y'/y)^2$, $B(y)=\frac{1}{y}\left(y''+2y'^2\frac{\omega'}{\omega}\right)$, and $C(y)=y'^2\frac{\omega''}{\omega}+y''\frac{\omega'}{\omega}$. Now, for the given basis $\left\{\phi_n(x)\right\}_{n=0}^{\infty}$, a solution of the system described by the wave equation (2.1) is obtained whence the wavefunction expansion coefficients $\left\{p_n(\varepsilon)\right\}_{n=0}^{\infty}$ are determined. Without significant loss of generality, we look for a class of solutions for which the coefficients $p_n(\varepsilon)$ satisfying a three-term recursion relation. However, it should be clearly stated that this, of course, does not necessarily mean that $p_n(\varepsilon)$ will be a polynomial of degree *n* in $\varepsilon$. It is only in special circumstances where this will be the case. We will continue to deal with the general case but limit our detailed investigation to situations where $\left\{p_n(\varepsilon)\right\}_{n=0}^{\infty}$ are, in fact, polynomials in $\varepsilon$ but do not necessarily form an orthogonal set. Now, Eq. (2.3) does give the sought after recursion relation if and only if (i) the functions *A*(*y*) and *B*(*y*) are linear sum in the monomials $y^{\sigma}$ and $y^{\sigma\pm\mu}$, where either $\sigma=0$ or $\sigma=\pm\mu$, and (ii) the potential function *v*(*y*) is chosen such that *C*(*y*) − *v*(*y*) is also proportional to $y^{\sigma}$ and $y^{\sigma\pm\mu}$. If we write,

$$A(y)=A_0y^{\sigma}+A_+y^{\sigma+\mu}+A_-y^{\sigma-\mu}\,, \tag{2.4a}$$

$$B(y)=B_0y^{\sigma}+B_+y^{\sigma+\mu}+B_-y^{\sigma-\mu}\,, \tag{2.4b}$$

$$C(y)-v(y)=C_0^vy^{\sigma}+C_+^vy^{\sigma+\mu}+C_-^vy^{\sigma-\mu}\,, \tag{2.4c}$$

where $\left\{A_0,A_\pm,B_0,B_\pm,C_0^v,C_\pm^v\right\}$ are real constant coefficients, then the resulting three-term recursion relation for $\sigma=0$ could be written as

$$\varepsilon p_n(\varepsilon)=a_np_n(\varepsilon)+d_n^-p_{n-1}(\varepsilon)+d_n^+p_{n+1}(\varepsilon)\,,\ n\geq 1 \tag{2.5}$$

where the recursion coefficients are

$$-a_n=\mu n\left[(\mu n-1)A_0+B_0\right]+C_0^v\,, \tag{2.6a}$$

$$-d_n^{\pm}=\mu(n\pm 1)\left[(\mu n\pm\mu-1)A_{\mp}+B_{\mp}\right]+C_{\mp}^v\,, \tag{2.6b}$$

and the initial relation (where, *n* = 0) is $\varepsilon p_0(\varepsilon)=a_0p_0(\varepsilon)+d_0^+p_1(\varepsilon)$. We choose a normalization in which $p_0:=1$ and refer to this case, where $\sigma=0$, as the "diagonal representation". On the other hand, for $\sigma=\pm\mu$ we obtain the following three-term recursion relation

$$C_0^vp_n(\varepsilon)=\tilde{a}_np_n(\varepsilon)+\left(d_n^-+\varepsilon\,\delta_-\right)p_{n-1}(\varepsilon)+\left(d_n^++\varepsilon\,\delta_+\right)p_{n+1}(\varepsilon)\,,\ n\geq 1 \tag{2.7}$$

where $\tilde{a}_n = a_n + C_0^v = -\mu n\left[(\mu n - 1)A_0 + B_0\right]$ and $\delta_\pm = \frac{\mu\pm\sigma}{2\mu}$, which is either 0 or 1. The initial relation for this case is $C_0^v p_0(\varepsilon) = \tilde{a}_0 p_0(\varepsilon) + \left(d_0^+ + \varepsilon\,\delta_+\right) p_1(\varepsilon)$ with $p_0 := 1$. This case will be referred to as the "off-diagonal representation".

It must be emphasized that the three-term recursion relations (2.5) and (2.7) are equivalent to the wave equation (2.1), $(H - E)|\psi_E\rangle = 0$. More precisely, they are the matrix representation of the wave equation in the complete $L^2$ basis, $\{\phi_n(x)\}_{n=0}^\infty$. A solution of the problem is completely determined whence the recursion relations (2.5) or (2.7) are solved for the expansion coefficients $\{p_n(\varepsilon)\}_{n=0}^\infty$. For example, if the recursion coefficients $\{a_n, d_n^\pm\}_{n=0}^\infty$ in (2.5) are known and the choice of normalization is made (e.g., $p_0 = 1$) then all $\{p_n(\varepsilon)\}_{n=0}^\infty$ are determined recursively as follows:

$$p_0 = 1,\; p_1 = (\varepsilon - a_0)/d_0^+\,, \tag{2.8a}$$

$$p_2 = \left(d_1^+\right)^{-1}\left[(\varepsilon - a_1)p_1 - d_1^-\right], \tag{2.8b}$$

$$p_3 = \left(d_2^+\right)^{-1}\left[(\varepsilon - a_2)p_2 - d_2^- p_1\right], \tag{2.8c}$$

……

$$p_n = \left(d_{n-1}^+\right)^{-1}\left[(\varepsilon - a_{n-1})p_{n-1} - d_{n-1}^- p_{n-2}\right]. \tag{2.8d}$$

……

Therefore, throughout this work we limit our investigation to these expansion coefficients. However, due to the infinite dimensionality of the problem, an exact solution is not guaranteed. Only when the problem is highly symmetric will such an exact solution be possible. Nonetheless, if one finds a natural scheme of making the problem finite then an exact partial solution is possible. In other words, if a method is devised whereby the recursion relations (2.5) and (2.7) terminates for the first *N* expansion coefficients, then all $\{p_n(\varepsilon)\}_{n=0}^{N-1}$ will be completely determined starting with $p_0 = 1$ giving an exact evaluation of the total wavefunction $\psi_\varepsilon(x)$. Moreover, under some reality constraints the energy spectrum $\{\varepsilon_n\}_{n=0}^{N-1}$ could be obtained exactly by several methods, one of which as the *N*-dimensional set of zeros of the polynomial $p_N(\varepsilon)$.

One can easily verify that the three-term recursion relation (2.5) for the diagonal representation with the initial value $p_0 := 1$, gives $p_n(\varepsilon)$ as a polynomial in $\varepsilon$ of degree *n*. On the other hand, relation (2.7) for the off-diagonal representation with $\sigma = -\mu$ gives $p_n(\varepsilon)$ as a polynomial of degree $\frac{n}{2}$ $\left(\text{degree } \frac{n-1}{2}\right)$ for even (odd) *n*. However, for $\sigma = +\mu$ $p_n(\varepsilon)$ will be a ratio of a polynomial of degree $\frac{n}{2}$ $\left(\text{degree } \frac{n-1}{2}\right)$ over a polynomial of degree *n* for even (odd) *n*, respectively. Nonetheless, in the off-diagonal representation ($\sigma = \pm\mu$) and for a fixed value of the energy, the recursion relation (2.7) could be thought of as describing an orthogonal polynomial set $\{p_n(\xi)\}$ in the "variable" $\xi = C_0^v$ that depends on the potential parameters. As such, several solutions are obtained at one special value of the energy but for a set of potential parameters (the parameter spectrum) that could be calculated by several methods, for example, from the zeros $\{\xi_n\}_{n=0}^{N-1}$ of $p_N(\xi)$. The advantage of this interpretation will become clear when we discuss problems related to this class in Sec. IV and Sec. V.

To give a clear and simpler physical interpretation, we make a transformation into a new complete set of $L^2$ basis $\{\chi_n(x)\}$ such that the corresponding expansion coefficients in $\psi_\varepsilon(x)=\sum_n q_n(\varepsilon)\chi_n(x)$ satisfy the following *symmetric* three-term recursion relation

$$\varepsilon q_n(\varepsilon)=a_n q_n(\varepsilon)+b_{n-1}q_{n-1}(\varepsilon)+b_n q_{n+1}(\varepsilon), \tag{2.5)'}$$

$$C_0^v q_n(\varepsilon)=\tilde{a}_n q_n(\varepsilon)+c_{n-1}q_{n-1}(\varepsilon)+c_n q_{n+1}(\varepsilon), \tag{2.7)'}$$

where it is assumed (in non-degenerate representations) that $b_n\neq 0$ and $c_n\neq 0$ for all $n$ [5]. These two relations could easily be obtained from (2.5) and (2.7) by the mapping $q_n(\varepsilon)=\Omega_n p_n(\varepsilon)$, which implies that $\Omega_0=1$ if we take $q_0=1$. Moreover, in this new representation the Hamiltonian matrix for the diagonal case (or the "Hamiltonian-equivalent" matrix for the off-diagonal case) becomes tridiagonal and symmetric (i.e., hermitian). This is the most significant property in the new realization given by (2.5)' and (2.7)'. It will guide our development and base it on sound physical principles. For the diagonal representation, the action of this map on (2.5) gives

$$b_n^2=d_n^+d_{n+1}^-,\ \Omega_{n+1}=\prod_{m=0}^n\left(d_m^+/d_{m+1}^-\right)^{\frac{1}{2}},\ \text{and}\ \chi_n(x)=\Omega_n^{-1}\phi_n(x). \tag{2.9}$$

On the other hand, for the off-diagonal representation we obtain

$$c_n^2(\varepsilon)=\begin{cases}\left(d_n^++\varepsilon\right)d_{n+1}^- & ,\sigma=+\mu\\ d_n^+\left(d_{n+1}^-+\varepsilon\right) & ,\sigma=-\mu\end{cases}, \tag{2.10a}$$

$$\Omega_{n+1}(\varepsilon)=\begin{cases}\prod_{m=0}^n\left[(d_m^++\varepsilon)/d_{m+1}^-\right]^{\frac{1}{2}} & ,\sigma=+\mu\\ \prod_{m=0}^n\left[d_m^+/(d_{m+1}^-+\varepsilon)\right]^{\frac{1}{2}} & ,\sigma=-\mu\end{cases} \tag{2.10b}$$

and $\chi_n(x,\varepsilon)=\Omega_n^{-1}(\varepsilon)\phi_n(x)$, which makes the new basis elements in the off-diagonal representation (except for $\chi_0$) energy-dependent. The three-term recursion relation (2.5)' could be written as $h|q\rangle=\varepsilon|q\rangle$, which is the matrix representation equivalent of the wave equation $(H-E)|\psi_E\rangle$ = 0 and with the following infinite tridiagonal symmetric Hamiltonian matrix

$$h=\begin{pmatrix} a_0 & b_0 & & & & & \\ b_0 & a_1 & b_1 & & \mathbf{0} & & \\ & b_1 & a_2 & b_2 & & & \\ & & b_2 & \times & \times & & \\ & & & \times & \times & \times & \\ & \mathbf{0} & & & \times & \times & \times \\ & & & & & \times & \times \end{pmatrix} \tag{2.11}$$

On the other hand, we can write the recursion relation (2.7)' as $\tilde{h}|q\rangle=C_0^v|q\rangle$, which is also equivalent to the wave equation but with the following "Hamiltonian-equivalent" matrix

$$\tilde{h}(\varepsilon)=\begin{pmatrix} \tilde{a}_0 & c_0(\varepsilon) & & & & & \\ c_0(\varepsilon) & \tilde{a}_1 & c_1(\varepsilon) & & \mathbf{0} & & \\ & c_1(\varepsilon) & \tilde{a}_2 & c_2(\varepsilon) & & & \\ & & c_2(\varepsilon) & \times & \times & & \\ & & & \times & \times & \times & \\ & \mathbf{0} & & & \times & \times & \times \\ & & & & & \times & \times \end{pmatrix} \tag{2.12}$$

The condition of existence of nontrivial solutions to the recursion relations is that $\det\left[\Im(\varepsilon)\right]=0$, where $\Im(\varepsilon)$ is the tridiagonal matrix operator defined by writing the recursion relations (2.5)' and (2.7)' in a matrix form as $\Im\left|q\right\rangle=0$. The solution of this equation (i.e., roots of $\left|\Im(\varepsilon)\right|$) is one of the methods for obtaining the energy spectrum $\left\{\varepsilon_n\right\}_{n=0}^{\infty}$.

As stated above, a given arbitrary set of parameters $\left\{A_0, A_\pm, B_0, B_\pm, C_0^v, C_\pm^v\right\}$ does not guarantee an exact solution of the three-term recursion relations (2.5) and (2.7) [or, equivalently, (2.5)' and (2.7)']. However if, for a given positive integer *N*, the condition that $b_{N-1}=0$ or $c_{N-1}(\varepsilon)=0$, for all *ε*, could be satisfied by these parameters then the matrix representation as given by Eqs. (2.11) and (2.12) will be completely reducible into two components. One of them is finite and associated with the *N*×*N* submatrix (top-left corner) of *h* or $\tilde{h}$ and the other is infinite and associated with the remaining infinitely long tail of these matrices. Under certain conditions the finite component becomes real and completely solvable whereas the infinite component might still be unsolvable. Therefore, the whole problem belongs to the quasi-exactly solvable class. In the following section we investigate this situation for the diagonal representation ($\sigma=0$) and give some examples. The off-diagonal representation will be discussed in Sec. IV and Sec. V for $\sigma=-\mu$ and $\sigma=+\mu$, respectively.

## III. THE DIAGONAL REPRESENTATION ($\sigma = 0$)

In this representation the three-term recursion relation (2.5) or (2.5)' results in a set of orthogonal polynomials in the energy *ε*. That is, there exist a positive measure $\rho(\varepsilon)$ such that $\int \rho(\varepsilon) q_n(\varepsilon) q_m(\varepsilon)\, d\varepsilon=\delta_{nm}$. Equivalently,

$$\int \rho(\varepsilon) p_n(\varepsilon) p_m(\varepsilon)\, d\varepsilon=\Omega_n^{-2}\delta_{nm}. \tag{3.1}$$

The condition $b_{N-1}=0$ is equivalent to either $d_{N-1}^{+}=0$ or $d_N^{-}=0$. This gives a constraint on the parameters $\left\{A_-, B_-, C_-^v\right\}$ or $\left\{A_+, B_+, C_+^v\right\}$, respectively, resulting in only special permissible values for some of the physical parameters for obtaining the quasi-exact solutions. Moreover, with $b_{N-1}=0$ the recursion relation (2.5)' results in all polynomials $q_{N+n}(\varepsilon)$ having the common factor $q_N(\varepsilon)$. That is, we obtain the factorization $q_{N+n}(\varepsilon)$ = $q_N(\varepsilon)\tilde{q}_n(\varepsilon)$. This could easily be verified by noting that the recursion relation (2.5)' with $n = N$ gives $q_{N+1}(\varepsilon)=[b_N^{-1}(\varepsilon-a_N)]q_N(\varepsilon)$. Thus, all subsequent polynomials will have $q_N(\varepsilon)$ as a common factor. This observation has already been reported elsewhere in the literature (see, for example, page 6 of [4]). Now, since $b_n$ does not vanish for all $n$ except when $n = N–1$ and since $b_n^2$ is real, then generally the condition $b_{N-1}=0$ makes $b_n^2$ change sign as $n$ jumps from $N–2$ to $N$. Consequently, if the parameters of the problem are chosen such that $b_n^2>0$ for $n\le N-2$ (or, $b_n^2<0$ for $n\ge N$), then $\{b_{n\le N-2}\}$ become real and $\{b_{n\ge N}\}$ become pure imaginary. Moreover, the matrix representation of the Hamiltonian shown in (2.11) becomes completely reducible as the direct sum $h=h_0^N\oplus h_1^N$, where $h_0^N$ is the finite $N\times N$ real symmetric tridiagonal matrix,

$$h_0^N=\begin{pmatrix} a_0 & b_0 & & & & & \\ b_0 & a_1 & b_1 & & \mathbf{0} & & \\ & b_1 & a_2 & b_2 & & & \\ & & b_2 & \times & \times & & \\ & & & \times & \times & \times & \\ & \mathbf{0} & & & \times & a_{N-2} & b_{N-2} \\ & & & & & b_{N-2} & a_{N-1} \end{pmatrix} \tag{3.2}$$

and $h_1^N$ is an infinite dimensional complex tridiagonal matrix,

$$h_1^N=\begin{pmatrix} a_N & b_N & & & & & \\ b_N & a_{N+1} & b_{N+1} & & \mathbf{0} & & \\ & b_{N+1} & a_{N+2} & b_{N+2} & & & \\ & & b_{N+2} & \times & \times & & \\ & & & \times & \times & \times & \\ & \mathbf{0} & & & \times & \times & \times \\ & & & & & \times & \times \end{pmatrix} \tag{3.3}$$

Consequently, the three-term recursion relation (2.5)’ splits into two disconnected and independent relations. The finite relation reads as follows:

$$\varepsilon\, q_n(\varepsilon)=a_n q_n(\varepsilon)+b_{n-1}q_{n-1}(\varepsilon)+b_n q_{n+1}(\varepsilon),\ N-1\ge n\ge 1, \tag{3.4}$$

with $q_0=1$ and initial relation $\varepsilon q_0(\varepsilon)=a_0q_0(\varepsilon)+b_0q_1(\varepsilon)$. Due to the fact that $b_{N-1}=0$, the end relation ($n$ = $N$–1) of (3.4) becomes $\varepsilon\, q_{N-1}(\varepsilon)=a_{N-1}q_{N-1}(\varepsilon)+b_{N-2}q_{N-2}(\varepsilon)$. The other infinite recursion relation could be written in terms of an independent set of polynomials $\hat{q}_n(\varepsilon):=q_{N+n}(\varepsilon)$ as

$$\mathrm{i}\varepsilon\, \hat{q}_n(\varepsilon)=\mathrm{i}\,\hat{a}_n\hat{q}_n(\varepsilon)+\hat{b}_{n-1}\hat{q}_{n-1}(\varepsilon)+\hat{b}_n\hat{q}_{n+1}(\varepsilon),\ n\ge 1 \tag{3.5}$$

where $\hat{a}_n=a_{n+N}$ and $\hat{b}_n=\mathrm{i}\,b_{N+n}$ are real. The initial relation is: $\mathrm{i}\varepsilon\,\hat{q}_0(\varepsilon)=\mathrm{i}\,\hat{a}_0\hat{q}_0(\varepsilon)$ + $\hat{b}_0\hat{q}_1(\varepsilon)$. Therefore, the real solution space of the problem collapses into a finite $N$-dimensional space. On the other hand, we now show that the infinite subspace has *zero norm* states. This statement could be verified by studying the relation between the two bases which is given as $q_n=\Omega_n p_n$ (or, $\chi_n=\Omega_n^{-1}\phi_n$). Now, if $d_{N-1}^{+}=0$ then the formula for $\Omega_n$ given in Eq. (2.9) shows that $\Omega_{n\ge N}=0$ making $q_{n\ge N}=0$, whereas if $d_N^{-}=0$ then $\Omega^{-1}_{n\geq N} = 0$ giving $\chi_{n\geq N} = 0$. Consequently, the infinite series expansion of the total wavefunction $\psi_\varepsilon(x)$ in the $\{\chi_n(x)\}$ basis truncates into only the first $N$ terms. Physically, this could also be understood by studying an illustrative model in which $A_0 = B_0 = 0$ and with a trivial constant shift in the energy by $-C_0^\nu$ (thus, $a_n = 0$). In such a model the matrix representation of the wave equation in the space spanned by $\{\chi_n\}_{n=N}^{\infty}$ becomes equivalent to the recursion relation $\mathrm{i}\varepsilon\,\hat{q}_n(\varepsilon) = \hat{b}_{n-1}\hat{q}_{n-1}(\varepsilon) + \hat{b}_n\hat{q}_{n+1}(\varepsilon)$. This implies that a real solution exists for pure imaginary energy (i.e., $\varepsilon \to -\mathrm{i}\,\varepsilon$). Therefore, the total time-dependent wavefunction decays in time due to the factor $e^{-\mathrm{i}\varepsilon t}$. Thus, the steady state solution of the time-independent Schrödinger equation (2.1) vanishes and what survive are only the finite $N$-dimensional solution space and its associated energy spectrum. This contraction of the physical solution space to a finite $N$-dimensional subspace will be explicitly demonstrated in the examples given below. Now, the energy spectrum $\{\varepsilon_n\}_{n=0}^{N-1}$ could be obtained by several equivalent methods. The following are three examples of such methods:

1) As the eigenvalues of the $N\times N$ real symmetric tridiagonal matrix $h_0^N$ of Eq. (3.2).
2) As the zeros of the determinant of the tridiagonal matrix operator $\Im(\varepsilon) = h_0^N - \varepsilon I$, where $I$ is the $N\times N$ unit matrix.
3) As the $N$ real roots of the polynomial $p_N(\varepsilon)$.

The last one provides another insight into the physical space contraction; at least for some special values of the energy. Due to the factorization $p_{N+n}(\varepsilon) = p_N(\varepsilon)\tilde{p}_n(\varepsilon)$ mentioned above, the expansion of the wavefunction at the energy eigenvalues $\{\varepsilon_n\}_{n=0}^{N-1}$ terminates after the first $N$ terms. That is, because $p_N(\varepsilon_n) = 0$, we can write $\psi_{\varepsilon_n}(x) = \sum_{n=0}^{\infty} p_n(\varepsilon_n)\phi_n(x) = \sum_{n=0}^{N-1} p_n(\varepsilon_n)\phi_n(x)$.

To illustrate our findings we present, as examples, two sets of quasi-exactly solvable problems and their solutions. In this section, we are interested only in the $\sigma = 0$ class. In the Appendix we give extra examples of potentials that belong to this diagonal representation subclass.

### A. Power Potentials

As stated above Eq. (2.4) our choice of coordinate transformation $y(x)$ and weight function $\omega(y)$ is restricted by the important constraints listed in Eq. (2.4). These are, of course, above and beyond the requirements of square integrability and compatibility with the boundary conditions (i.e., compatibility with the domain of the Hamiltonian). As a simple illustration, we start by reproducing the Bender-Dunne potential model [4] where $y(x) = x$ and $\omega(y) = y^\gamma e^{-\alpha y^\beta}$. Square integrability and the boundary conditions require that $\alpha$, $\beta$ and $\gamma$ be real and positive. Moreover, we obtain the following functions in Eqs. (2.4)

$$A(y) = y^{-2},\ B(y) = 2y^{-2}\left(\gamma - \alpha\beta y^\beta\right),\ \text{and} \tag{3.6a}$$

$$C(y) = y^{-2}\left[\gamma(\gamma-1) - \alpha\beta(2\gamma+\beta-1)y^\beta + \alpha^2\beta^2 y^{2\beta}\right]. \tag{3.6b}$$

Therefore, due to the fact that $A(y)$ and $B(y)$ are sums of terms proportional to $y^{-2}$ and $y^{\beta-2}$, then we end up with three possibilities for Eqs. (2.4): (1) $\sigma=-2$ and $\mu=\beta$, (2) $\sigma=\beta-2$ and $\mu=\beta$, or (3) $\sigma=\frac{1}{2}\beta-2$ and $\mu=\frac{1}{2}\beta$. Moreover, the potential function $v(y)$ is chosen such that $C(y) - v(y)$ must also be a sum of terms that are only proportional to $y^{\sigma}$ and $y^{\sigma\pm\mu}$. Thus, we obtain the following three corresponding potential functions

$$(1)\ \sigma=-2:\ v(y)=\alpha^2\beta^2 y^{2(\beta-1)}+v_1\,y^{-2}+v_2\,y^{\beta-2}+v_3\,y^{-\beta-2}, \tag{3.7a}$$

$$(2)\ \sigma=\beta-2:\ v(y)=v_1\,y^{-2}+v_2\,y^{\beta-2}+v_3\,y^{2(\beta-1)}, \tag{3.7b}$$

$$(3)\ \sigma=\tfrac{1}{2}\beta-2:\ v(y)=\alpha^2\beta^2 y^{2(\beta-1)}+v_1\,y^{-2}+v_2\,y^{\beta-2}+v_3\,y^{\frac{1}{2}\beta-2}. \tag{3.7c}$$

where $v_1$, $v_2$, and $v_3$ are real potential parameters that are, at present, arbitrary but will be restricted as we go. The diagonal representation (where $\sigma=0$) is obtained only in the second and third cases with $\beta$ = 2 and $\beta$ = 4, respectively. The second case, where $v(x)=v_1\,x^{-2}+v_3\,x^2$, corresponds to the 1D harmonic oscillator problem plus an inverse square potential barrier or to the 3D isotropic oscillator with an orbital term where $v_1=\ell(\ell+1)$. However, the third case corresponds to the Bender-Dunne model [4] where

$$v(x)=16\alpha^2x^6+v_1\,x^{-2}+v_2\,x^2, \tag{3.8a}$$

$$\phi_n(x)=x^{\gamma}e^{-\alpha x^4}x^{2n}. \tag{3.8b}$$

The constant $v_3$ term in the potential was absorbed in the energy. Now, Eqs. (3.6) together with Eq. (3.8a) give $A_0=A_+=B_0=C_0^v=0$, $A_-=1$, $B_+=-8\alpha$, $B_-=2\gamma$, $C_+^v$ = $-8\alpha\left(\gamma+3/2\right)-v_2$, and $C_-^v=\gamma(\gamma-1)-v_1$. Substituting these in Eqs. (2.6) gives the coefficients for the three-term recursion relation (2.5) as $a_n=0$, $d_n^+=-\left(2n+\gamma+3/2\right)^2+v_1+\frac{1}{4}$, and $d_n^-=8\alpha\left(2n+\gamma-1/2\right)+v_2$. To achieve quasi-exact solvability ($b_{N-1}=0$) we can either (a) take $d_N^-=0$ by choosing the potential parameter $v_2=-8\alpha\left(2N+\gamma-1/2\right)$, or (b) take $d_{N-1}^+=0$ by choosing the potential parameter $v_1=\left(2N+\gamma\right)\left(2N+\gamma-1\right)$. The corresponding recursion relations become

$$\text{(a)}\ \varepsilon p_n(\varepsilon)=-16\alpha\left(N-n\right)p_{n-1}(\varepsilon)-\left[\left(2n+\gamma+3/2\right)^2-v_1-\tfrac{1}{4}\right]p_{n+1}(\varepsilon), \tag{3.9a}$$

$$\text{(b)}\ \varepsilon p_n(\varepsilon)=\left[8\alpha\left(2n+\gamma-1/2\right)+v_2\right]p_{n-1}(\varepsilon)$$
$$+4(N-n-1)\left(N+n+\gamma+1/2\right)p_{n+1}(\varepsilon), \tag{3.9b}$$

where $n=1,2,..,N-1$. The parameters in [4] correspond to case (a) with $\alpha=\frac{1}{4}$, $N = J$, $v_1=\gamma(\gamma-1)$, and $\gamma=2s-\frac{1}{2}$, where $s$ is real and $J=1,2,...$ Therefore, the resulting three-term recursion relation (3.9a) reads as follows

$$\varepsilon p_n(\varepsilon)=-4(J-n)p_{n-1}(\varepsilon)-4(n+1)(n+2s)p_{n+1}(\varepsilon),\ J-2\geq n\geq 1 \tag{3.10}$$

which is identical to the recursion relation of Eq. (6) on page 7 of [4] for the Bender-Dunne polynomial $[(-4)^n n!\,\Gamma(n+2s)]\,p_n(\varepsilon)$. Using the values of $d_n^+$ and $d_n^-$ one can evaluate $\Omega_n$ for this problem as shown in Eq. (2.9). Thus, we can write the orthogonality relation (3.1) for these two-parameter polynomials that satisfy (3.10) as

$$\int\rho(\varepsilon)p_n(\varepsilon)p_m(\varepsilon)\,d\varepsilon=\frac{\Gamma(J)\,\Gamma(2s)}{n!\,\Gamma(J-n)\Gamma(n+2s)}\delta_{nm}, \tag{3.11}$$

which is identical to Eq. (12) in Ref. [4]. Now, due to the gamma function, $\Gamma(J-n)$, in the denominator, the norm of $p_n(\varepsilon)$ vanishes for all $n \geq J$. Consequently, the expansion $\psi_\varepsilon = \sum_n p_n \phi_n$ terminates to the first $J$ terms resulting in a normalizable wavefunction. This is a verification example of the statement made below Eq. (3.5) that the physical solution space for quasi-exactly solvable problems collapses into a finite $N$-dimensional subspace whereas the remaining infinite one has zero norm states. Further details about this system could be found in [4].

One should also make note of the new quasi-exact solution, which is associated with case (b) for the related potential model

$$v(x) = 16\alpha^2 x^6 + (2N+\gamma)(2N+\gamma-1)x^{-2} + v_2\, x^2, \qquad (3.12)$$

where $v_2$ is a continuous parameter restricted by the reality of the representation ($b_n^2 > 0$ for all $n \leq N-2$) to be larger than the critical value $\tilde{v} = -4\alpha(2\gamma+3)$. For positive values it could be interpreted as the square of the oscillator frequency. Interpreting the second term in the potential (3.12) as the orbital $\ell(\ell+1)/x^2$, we conclude that the choice for $N$ is restricted by $N < \frac{1}{2}(\ell+1)$ so that $\gamma$ maintains positivity. Writing $v_2$ as $\tilde{v} + 16\alpha\,\xi$ with $\xi > 0$ and then substituting the values of $d_n^+$ and $d_n^-$ from (3.9b) into Eq. (2.9) we obtain an expression for $\Omega_n$ which when used in (3.1) gives the following orthogonality relation for the energy polynomials of this problem

$$\int \rho(\varepsilon) p_n(\varepsilon) p_m(\varepsilon)\, d\varepsilon = (4\alpha)^n \frac{\Gamma(N-n)\Gamma(N+\gamma+1/2)\Gamma(n+\xi)}{\Gamma(N)\Gamma(N+n+\gamma+1/2)\Gamma(\xi)} \delta_{nm}\,;\; n,m \leq N-1 \qquad (3.13)$$

For a given integer $N$ the energy spectrum, $\{\varepsilon_n\}_{n=0}^{N-1}$, could be obtained as the roots of the characteristic polynomial $\left|h_0^N - \varepsilon I\right|$ (i.e. solution of $\det\left[\Im(\varepsilon)\right] = 0$). For example, for $N = 2$ and $N = 3$ we obtain $\varepsilon = \pm 8\sqrt{\alpha\,\xi\left(\gamma+5/2\right)}$ and $\varepsilon = 0, \pm 8\sqrt{\alpha}\sqrt{\xi\left(3\gamma+23/2\right)+\gamma+9/2}$, respectively. Further analysis of these polynomials, their weight function and associated moments, etc. will not be pursued here but will be left for an appropriate mathematical setting. Nonetheless, we give here a sample graphical illustration. Figure 1 shows the weight function $\rho(\varepsilon)$ for $N = 10$ and for a given set of values of the physical parameters. The "Analytic Continuation" method developed in [6] was used in the evaluation of this weight function. Figure 2 is a plot of the first few polynomials where we also show the corresponding energy spectrum for $N = 7$.

## B. Exponential Potentials

As a second example, we take $y(x) = e^{-x}$ and the weight function is the same as above, $\omega(y) = y^\gamma e^{-\alpha y^\beta}$. Consequently, we obtain the following functions in Eqs. (2.4)

$$A(y) = 1,\; B(y) = 2\gamma + 1 - 2\alpha\beta y^\beta, \text{ and} \qquad (3.14a)$$

$$C(y) = \gamma^2 + \alpha^2\beta^2 y^{2\beta} - \alpha\beta(2\gamma+\beta) y^\beta. \qquad (3.14b)$$

Therefore, $A(y)$ and $B(y)$ are sums of the monomials $y^0$ and $y^\beta$. Thus, we end up again with three possibilities: (1) $\sigma = 0$ and $\mu = \beta$, (2) $\sigma = \mu = \beta$, or (3) $\sigma = \mu = \frac{1}{2}\beta$. Moreover, the potential function $v(y)$ is chosen such that $C(y) - v(y)$ must be a sum of the monomials $y^{\sigma}$ and $y^{\sigma\pm\mu}$ only. Therefore, we obtain the following three associated potentials

(1) $\sigma=0$: $v(y)=\alpha^2\beta^2 y^{2\beta}+v_1\, y^{\beta}+v_2\, y^{-\beta}$, (3.15a)

(2) $\sigma=\beta$: $v(y)=v_1\, y^{\beta}+v_2\, y^{2\beta}$, (3.15b)

(3) $\sigma=\frac{1}{2}\beta$: $v(y)=\alpha^2\beta^2 y^{2\beta}+v_1\, y^{\beta}+v_2\, y^{\frac{1}{2}\beta}$. (3.15c)

where $v_1$ and $v_2$ are real parameters. The second case corresponds to the Morse potential [7]. The third case is a generalization of the Morse potential $\alpha^2\beta^2 e^{-2\beta x}+v_1\, e^{-\beta x}$ (the decaying exponential and its square) that includes the square root of the exponential $e^{-\frac{1}{2}\beta x}$. However, these two cases belong to the off-diagonal $\sigma=+\mu$ class which will be discussed in Sec. V. The diagonal case is only the first one for which the potential function reads as follows

$$v(x)=\alpha^2\beta^2 e^{-2\beta x}+v_1\, e^{-\beta x}+v_2\, e^{\beta x}, \tag{3.16}$$

which is a combination of the Morse potential and an exponentially rising term. Now, Eqs. (3.14) together with Eq. (3.16) give $A_0=1$, $A_\pm=0$, $B_0=2\gamma+1$, $B_+=-2\alpha\beta$, $B_-=0$, $C_0^v=\gamma^2$, $C_+^v=-\alpha\beta(2\gamma+\beta)-v_1$, and $C_-^v=-v_2$. Inserting these values in Eqs. (2.6) gives the coefficients for the three-term recursion relation (2.5) as $a_n=-(\beta n+\gamma)^2$, $d_n^+=v_2$, and $d_n^-=2\alpha\beta^2\left(n-\frac{1}{2}+\gamma/\beta\right)+v_1$. To achieve quasi-exact solvability ($b_{N-1}=0$) we should either make $d_{N-1}^+=0$ by choosing $v_2=0$, which reduces the problem to the Morse potential, or make $d_N^-=0$ by choosing the potential parameter $v_1$ as

$$v_1=-2\alpha\beta^2\left(N-\tfrac{1}{2}+\gamma/\beta\right), \tag{3.17}$$

with $N=1,2,3,...$ Adapting the latter choice leads to the following recursion relation

$$\varepsilon p_n(\varepsilon)=-(\beta n+\gamma)^2 p_n(\varepsilon)-2\alpha\beta^2(N-n)p_{n-1}(\varepsilon)+v_2\, p_{n+1}(\varepsilon), \tag{3.18}$$

where $n=1,2,..,N-1$. Choosing $v_2<0$ makes $\{b_{n\le N-2}\}$ real and the problem becomes quasi-exactly solvable as explained above. We can simplify (3.18) by rescaling length (i.e., $\lambda$) and the parameters of the problem as follows:

$$x\to\beta^{-1}x,\ \varepsilon\to\beta^2\varepsilon,\ \gamma\to\beta\gamma, \text{ and } v_2\to\beta^2 v_2, \tag{3.19}$$

which is equivalent to making $\beta=1$. As a result, the potential function will take the following form

$$v(x)=\alpha^2 e^{-2x}-2\alpha\left(N+\gamma-1/2\right)e^{-x}-\xi\, e^{+x}, \tag{3.16'}$$

where we have written $v_2$ as the negative of a real parameter $\xi>0$. Moreover, the recursion relation (3.18) will be recast as follows

$$\varepsilon p_n(\varepsilon)=-(n+\gamma)^2 p_n(\varepsilon)-2\alpha(N-n)p_{n-1}(\varepsilon)-\xi p_{n+1}(\varepsilon), \tag{3.18'}$$

Substituting the values of $d_n^+$ and $d_n^-$ in the product formula for $\Omega_n$ in Eq. (2.9) we can write the orthogonality relation for these polynomials as

$$\int\rho(\varepsilon)p_n(\varepsilon)p_m(\varepsilon)d\varepsilon=\left(2\alpha/\xi\right)^n\frac{\Gamma(N)}{\Gamma(N-n)}\delta_{nm}. \tag{3.20}$$

This is another example showing that the physical solution space collapses into a finite $N$-dimensional subspace since the remaining infinite one has zero norm states as evidenced by the vanishing of the norm of the expansion coefficients, $p_n(\varepsilon)$, for all $n\ge N$. Now, the energy spectrum $\{\varepsilon_n\}_{n=0}^{N-1}$ could be obtained by any one of the three methods stated above.

For example, taking $N = 1$, we obtain $\varepsilon = -\gamma^2$, whereas for $N = 2$ the two eigen-energies are $\varepsilon = -\frac{1}{2} - \gamma(\gamma+1) \pm \sqrt{(\gamma+1/2)^2 + 2\alpha\xi}$ . Analysis of the properties of the polynomials satisfying (3.18)' will not be carried out here but will be pursued in another more appropriate future setting. Nevertheless, we give in Fig. 3 a plot of the weight function $\rho(\varepsilon)$ for $N = 20$ and for a given set of values of the potential parameters. Figure 4 shows the first few polynomials and the corresponding energy spectrum for $N = 7$. Next, we turn attention to the off-diagonal representation where $\sigma = \pm\mu$ .

## IV. THE OFF-DIAGONAL REPRESENTATION ($\sigma = -\mu$)

As stated in Sec. II, the energy polynomials $p_n(\varepsilon)$ in this representation are of degree $\frac{n}{2}$ $\left(\text{degree } \frac{n-1}{2}\right)$ for even (odd) $n$. The three-term recursion relation (2.7) in this case with $\sigma = -\mu$ reads as follows

$$C_0^v p_n(\varepsilon) = \tilde{a}_n p_n(\varepsilon) + \left(d_n^- + \varepsilon\right) p_{n-1}(\varepsilon) + d_n^+ p_{n+1}(\varepsilon) . \tag{4.1}$$

One is to observe the curious appearance of the polynomial variable $\varepsilon$ in the factor multiplying $p_{n-1}$ rather than $p_n$. This property together with the normalization $p_0 = 1$ is, in fact, the reason behind the behavior of the degree of these polynomials. Additionally, these polynomials can not form an orthogonal set with respect to any *positive* measure. One can proof this assertion by a counter example as follows. If one assumes that such positive measure, $\rho(\varepsilon)$, exists then we could write $\int \rho(\varepsilon) p_n(\varepsilon) p_m(\varepsilon)\, d\varepsilon \sim \delta_{nm}$. Now, taking $n = 0$ and $m = 1$ shows that $\int \rho(\varepsilon)\, d\varepsilon = 0$ since $p_0(\varepsilon)$ and $p_1(\varepsilon)$ are constants. Thus, $\rho(\varepsilon)$ can not be positive definite. Now, according to the General Formulation above, quasi-exact solvability condition in the off-diagonal representation is $c_{N-1}(\varepsilon) = 0$ for all $\varepsilon$. In the $\sigma = -\mu$ case this condition is equivalent to either $d_{N-1}^+ = 0$ or $d_N^- + \varepsilon = 0$. The first one restricts some of the physical parameters in the set $\left\{A_-, B_-, C_-^v\right\}$ to $N$-dependent fixed values. The second gives a solution of the problem at only one value of the energy $\varepsilon = \varepsilon_N = -d_N^-$; other solutions are obtained by varying the value of $N$ over the desired range in $N = 1, 2, \ldots$ Moreover, a real finite representation is obtained only if $c_n^2(\varepsilon)$ is positive for all $n \le N - 2$. This is an energy-dependent constraint that might not allow for physical solutions below or above some energy threshold, $\hat{\varepsilon}$. Due to this highly nontrivial property, one may not be able to make further general observations about problems in this class but ought to study each one individually. Now, if all conditions of quasi-exact solvability are satisfied then the Hamiltonian-equivalent matrix shown in (2.12) becomes completely reducible as the direct sum $\tilde{h} = \tilde{h}_0^N \oplus \tilde{h}_1^N$ . $\tilde{h}_0^N$ is a finite $N\times N$ real symmetric tridiagonal matrix, whereas $\tilde{h}_1^N$ an infinite dimensional complex matrix. Now, the condition of existence of nontrivial solutions to the recursion relation (4.1) is $\det\left[\Im(\varepsilon)\right] = 0$, where $\Im(\varepsilon) = \tilde{h}_0^N(\varepsilon) - C_0^v I$ and $\Im\left|q\right\rangle = 0$. This condition gives the energy spectrum $\left\{\varepsilon_n\right\}_{n=0}^{N-1}$ for the case when the solvability requirement is $d_{N-1}^+ = 0$. However, for the alternative solvability requirement $\varepsilon = \varepsilon_N = -d_N^-$, $\det\left[\Im(\varepsilon_N)\right] = 0$ gives in principle $N$ possible values for the physical parameters that are compatible with the solution of the problem at the energy $\varepsilon=\varepsilon_N$. These parameter values are elements of the set that we refer to as the "parameter spectrum".

We limit the illustrative examples to those in subsection A above where $\phi_n(x)$ = $x^\gamma e^{-\alpha x^\beta} x^{\mu n}$ and with all parameters real and positive. Another non-exhaustive list of potentials in this subclass is given in the Appendix. The quasi-exactly solvable potentials that belong to this subclass ($\sigma=-\mu$) are obtained from Eqs. (3.7a-3.7c) as follows:

(1) $\sigma=-2$, $\beta=2$: $v(x)=4\alpha^2x^2+v_1x^{-2}+v_3x^{-4}$. (4.2a)

(2) $\sigma=-1$, $\beta=1$: $v(x)=v_1x^{-2}+v_2x^{-1}$. (4.2b)

(3) $\sigma=-1$, $\beta=2$: $v(x)=4\alpha^2x^2+v_1x^{-2}+v_3x^{-1}$. (4.2c)

Case (2) corresponds either to the Coulomb problem in 3D with non-zero orbital term or to the 1D hydrogen atom [8] with an inverse square potential barrier. This is an exactly solvable problem. Of course, all exactly-solvable problems are quasi-exactly solvable, but the reverse is not true. Now, cases (1) and (3) are highly significant and interesting. Case (1) is for the oscillator potential plus the highly singular inverse-quartic potential [9]. In this case and if we write $v_1=\ell(\ell+1)$, where $\ell$ is the angular momentum quantum number, then the coefficients for the recursion relation (4.1) are: $C_0^v=-(\ell+\gamma)(\ell-\gamma+1)$, $\tilde{a}_n$ = $-4n(n+\gamma-1/2)$, $d_n^+=v_3$, and $d_n^-=4\alpha(2n+\gamma-3/2)$ giving

$$(\ell+\gamma)(\ell-\gamma+1)p_n(\varepsilon)=4n(n+\gamma-1/2)\,p_n(\varepsilon)$$
$$-\left[4\alpha(2n+\gamma-3/2)+\varepsilon\right]p_{n-1}(\varepsilon)-v_3\,p_{n+1}(\varepsilon). \quad (4.3)$$

However, the quasi-exact solvability condition, $c_{N-1}(\varepsilon)=0$ has two consequences. It either requires $v_3=0$ or $\varepsilon=\varepsilon_N=-4\alpha(2N+\gamma-3/2)$. The first choice reduces (4.3) to a two-term recursion relation of the harmonic oscillator problem, which is well-known and exactly solvable [10]. The second makes the problem (quasi-exactly) solvable at only one value of the energy, $\varepsilon_N$, but for several values of the potential parameter $v_3$ that must satisfy the equation $\det\left[\Im(\varepsilon_N)\right]=0$. To obtain the larger set of solutions one must vary the value of the positive integer *N*. Now, it should be obvious that $\det\left[\Im(\varepsilon_N)\right]=0$ gives a maximum of *N* permissible values for $v_3$ (the parameter spectrum). However, a real representation, in the subspace $\{\chi_n\}_{n=0}^{N-1}$, is achievable [i.e., $c_n^2(\varepsilon_N)>0$ for $n\le N-1$] only for negative values of $v_3$. Therefore, if we write $v_3$ in terms of a real parameter $\xi$ as $v_3=-2\alpha\,\xi^2$ then we can recast relation (4.3) with $\varepsilon=\varepsilon_N$ in terms of the polynomials $\{q_n\}$ defined in Sec. II as follows

$$\frac{1}{4\alpha\xi}(2n+\gamma+\ell)(2n+\gamma-\ell-1)q_n(\xi)=\sqrt{N-n}\,q_{n-1}(\xi)+\sqrt{N-n-1}\,q_{n+1}(\xi), \quad (4.4)$$

If we choose $\gamma=\ell+1$ then analysis of this recursion relation shows that $\det(\Im)=0$ results in symmetric $\xi$ roots, one of them is always $\xi$ = 0 with multiplicity two. It also gives *N* – 2 (*N* – 3) nonzero roots for even (odd) *N*, respectively. As examples, for *N* = 4, *N* = 5 we obtain $\alpha\xi=\pm\sqrt{6\left(\ell+\frac{5}{2}\right)\left(\ell+\frac{7}{2}\right)}$, $\alpha\xi=\pm2\sqrt{3\left(\ell+\frac{5}{2}\right)\left(\ell+\frac{7}{2}\right)\left(\ell+\frac{9}{2}\right)/\left(5\ell+\frac{41}{2}\right)}$, respectively. These two cases constitute the lowest nontrivial solutions for the potential (4.2a) with $v_1=\ell(\ell+1)$, $v_3=-2\alpha\,\xi^2$ and at the energies $\varepsilon=-4\alpha\left(\ell+15/2\right)$ and $\varepsilon=-4\alpha\left(\ell+19/2\right)$, respectively.

Case (3) is uniquely significant due to the fact that it is a new attempt (among very few) to obtain exact solutions to this interesting problem that combines the two most popular potentials in quantum mechanics; the oscillator (with frequency $2\alpha$) and Coulomb (with charge $v_3$) [11,12]. Using Eqs. (3.6) and Eq. (4.2c) we obtain the following recursion coefficients $C_0^v=-v_3$, $\tilde{a}_n=0$, $d_n^+=-\left(n+\gamma+1/2\right)^2+\frac{1}{4}+v_1$, and $d_n^-$ = $4\alpha\left(n+\gamma-1/2\right)$ giving

$$v_3\,p_n(\varepsilon)=-\left[4\alpha\left(n+\gamma-1/2\right)+\varepsilon\right]p_{n-1}(\varepsilon)+\left[\left(n+\gamma+1/2\right)^2-\tfrac{1}{4}-v_1\right]p_{n+1}(\varepsilon)\,. \quad (4.5)$$

This is identical to the recursion relation obtained for the same problem by Alberg and Wilets in [12] as Eq. (12) on page 9 with $v_1=\ell(\ell+1)$ (in the notation of [12] $v_3=2\lambda$, $\gamma=\ell+1$, $\alpha=c/2$, and $\varepsilon=-2\epsilon$). Now, one of the two quasi-exact solvability conditions ($d_{N-1}^+=0$) dictates that $v_1=(N+\gamma)(N+\gamma-1)$ and maps (4.5) into

$$v_3\,p_n(\varepsilon)=-\left[4\alpha\left(n+\gamma-1/2\right)+\varepsilon\right]p_{n-1}(\varepsilon)-(N-n-1)(N+n+2\gamma)p_{n+1}(\varepsilon)\,. \quad (4.6)$$

Moreover, reality of the representation [i.e., $c_n^2(\varepsilon)$ is positive for all $n\le N-2$] requires that the energy be above the threshold $\hat{\varepsilon}=-2\alpha(2\gamma+1)$. The energy spectrum could be obtained as solutions of $\det\left[\Im(\varepsilon)\right]=0$. For example, when $N$ = 2, we obtain the energy eigenvalue $\varepsilon=-2\alpha(2\gamma+1)+v_3^2/2(\gamma+1)$. On the other hand, if $N$ = 3 then $\varepsilon$ = $-\dfrac{6\alpha}{3\gamma+5}(\gamma+1)(2\gamma+3)+\dfrac{v_3^2/2}{3\gamma+5}$. Now, the other solvability condition ($d_N^-+\varepsilon=0$) shows that a quasi-exact solution is obtained at the energy $\varepsilon=\varepsilon_N=-4\alpha\left(N+\gamma-1/2\right)$, which is Eq. (13) in Ref. [12]. But in this case the permissible values for the potential parameter $v_3$ must satisfy the equation $\det\left[\Im(\varepsilon_N)\right]$ = 0. Writing $v_1=\ell(\ell+1)$ and repeating the same analysis that was done above for the potential (4.2a) with $\varepsilon=\varepsilon_N$ leads to the following recursion relation in terms of the polynomials $\{q_n\}$

$$\begin{aligned} v_3\,q_n(v_3)&=2\sqrt{\alpha(N-n)(n+\gamma+\ell)(n+\gamma-\ell-1)}\,q_{n-1}(v_3)\\ &\quad+2\sqrt{\alpha(N-n-1)(n+\gamma+\ell+1)(n+\gamma-\ell)}\,q_{n+1}(v_3)\,. \end{aligned} \quad (4.7)$$

Reality of the representation requires that $\gamma\ge\ell+1$. Moreover, $\det\left(\Im\right)=0$ for this recursion relation produces the parameter spectrum which consists of $N$ real values for the potential parameter $v_3$ symmetrically distributed around $v_3$ = 0. For $N$ = 2 and $N$ = 3, where the energies are $\varepsilon=\varepsilon_2=-4\alpha\left(\gamma+3/2\right)$ and $\varepsilon=\varepsilon_3=-4\alpha\left(\gamma+5/2\right)$, these values are $v_3$ = $\pm2\sqrt{\alpha(\gamma+\ell+1)(\gamma-\ell)}$ and $v_3=0,\pm2\alpha^{\frac{1}{2}}\sqrt{\gamma(3\gamma+5)-3\ell(\ell+1)+2}$, respectively. For further analysis and details of the solution of the problem the reader may consult Ref. [12]. Next, we investigate the subclass of the off-diagonal representation where $\sigma=+\mu$.

## V. THE OFF-DIAGONAL REPRESENTATION ($\sigma = +\mu$)

The expansion coefficients of the wavefunction in this subclass satisfies the following three-term recursion relation

$$C_0^v p_n(\varepsilon) = \tilde{a}_n p_n(\varepsilon) + d_n^- p_{n-1}(\varepsilon) + \left(d_n^+ + \varepsilon\right) p_{n+1}(\varepsilon)\,,\; n \geq 1\,. \tag{5.1}$$

The appearance of the recursion variable $\varepsilon$ in the factor multiplying $p_{n+1}$ rather than $p_n$ together with the normalization $p_0 = 1$ is the reason that these are not polynomials but ratios of polynomials. However, if the quasi-exact solvability condition is satisfied and the normalization is changed form $p_0 = 1$ to $p_{N-1} = 1$, then we can reverse the recursion process in (5.1) by rewriting it as follows

$$p_n(\varepsilon) = -\left(d_{n+1}^-\right)^{-1}\left[a_{n+1}p_{n+1}(\varepsilon) + \left(d_{n+1}^+ + \varepsilon\right)p_{n+2}(\varepsilon)\right],\; n = N-2, N-3,..,1\,. \tag{5.2}$$

and with $p_N \equiv 0$. This relation should be supplemented by the initial relation of (5.1) that reads $p_0(\varepsilon) = -[(d_0^+ + \varepsilon)/a_0]p_1(\varepsilon)$, which is the final relation for (5.2). Consequently, the polynomials $p_n(\varepsilon)$ in this representation with the choice of normalization $p_{N-1} = 1$, are of degree $\frac{N-n-1}{2}$ $\left(\text{degree } \frac{N-n-2}{2}\right)$ for odd (even) $N$–$n$, where $n = 0,1,..,N-1$. One should observe that the choice of normalization in (5.1) is arbitrary but has to be fixed once and for all. This is due to the fact that a solution of (5.1) is unique modulo an arbitrary nonzero function of $\varepsilon$ that is independent of *n*. We took $p_{N-1}(\varepsilon)$ as the arbitrary function. One can also show that in this subclass and similar to the previous one, where $\sigma = -\mu$, these polynomials can not form an orthogonal set with respect to any *positive* energy measure. Moreover, the quasi-exact solvability condition $d_N^- = 0$ restricts some of the physical parameters in the set $\left\{A_+, B_+, C_+^v\right\}$ to only discrete values. The energy spectrum could be obtained by the requirement $\det\left[\Im(\varepsilon)\right] = 0$, where again $\Im(\varepsilon) = \tilde{h}_0^N(\varepsilon) - C_0^v I$ and $\Im\left|q\right\rangle = 0$. On the other hand, the alternative solvability condition, $d_{N-1}^+ + \varepsilon = 0$, gives a solution at $\varepsilon = \varepsilon_N = -d_{N-1}^+$ while restricting the problem parameters to satisfy the constraint $\det\left[\Im(\varepsilon_N)\right] = 0$ (the parameter spectrum). To obtain a larger set of solutions, one varies the value of the integer *N*. The real representation requirement [that is, $c_n^2(\varepsilon)$ is positive for all $n \leq N-2$] could force an energy threshold on the physical solutions. Moreover, complete reducibility of the representation and reduction of the solution space is similar to that in the previous section. Illustrative examples will be limited to those problems presented in subsection B of Sec. III where $y(x) = e^{-x}$. Other examples in this subclass are given in the Appendix.

Among the three potentials in subsection B above, the only interesting one which belongs to the off-diagonal representation $\sigma = +\mu$ is (3.15c) with $\sigma = \mu = \frac{1}{2}\beta$ and for the following generalization of the Morse potential

$$v(x) = \alpha^2\beta^2 e^{-2\beta x} + v_1\, e^{-\beta x} + v_2\, e^{-\frac{1}{2}\beta x}\,. \tag{5.3}$$

Nonetheless, the potential (3.15b) does belong to the subclass $\sigma = +\mu$ but is not of interest to our study since it corresponds to the exactly solvable *S*-wave Morse potential. Now, rescaling length and the problem parameters as done in (3.19) will have the equivalent effect of choosing $\beta = 1$. Thus, the coefficients of the recursion relation (5.1) associated with the potential (5.3) are: $C_0^v = -v_2$, $\tilde{a}_n = 0$, $d_n^+ = -\frac{1}{4}(n+2\gamma+1)^2$, and $d_n^- = \alpha(n+2\gamma)+v_1$. Moreover, the quasi-exact solvability condition $d_N^- = 0$ requires that $v_1 = -\alpha(N+2\gamma)$ giving

$$v_2\, p_n(\varepsilon) = \alpha(N-n)p_{n-1}(\varepsilon) + \left[\tfrac{1}{4}(n+2\gamma+1)^2 - \varepsilon\right]p_{n+1}(\varepsilon)\,. \tag{5.4}$$

Requiring reality of the representation [that is, $c_n^2(\varepsilon)$ is positive for all $n \le N-2$] dictates that the energy should be below the threshold $\hat{\varepsilon} = (\gamma+1/2)^2$. As stated above, the energy spectrum could be obtained in this case as solution of the characteristic equation $\det[\Im(\varepsilon)] = 0$. For example, with $N = 2$ we obtain $\varepsilon = (\gamma+1/2)^2 - v_2^2/\alpha$, whereas for $N = 3$ the result is $\varepsilon = \frac{2}{3}(\gamma+1/2)^2 + \frac{1}{3}(\gamma+1)^2 - v_2^2/3\alpha$. On the other hand, the alternative solvability constraint $d_{N-1}^+ + \varepsilon = 0$ gives a solution of the problem at the energy $\varepsilon = \varepsilon_N = (\gamma+N/2)^2$. In this case, the expansion coefficients of the wavefunction at this energy could be thought of as polynomials in the potential parameter $v_2$ of degree $n$ and satisfying the three-term recursion relation

$$-v_2\, p_n(v_2) = \left[\alpha(n+2\gamma)+v_1\right]p_{n-1}(v_2) + \tfrac{1}{4}(N-n-1)(N+n+4\gamma+1)p_{n+1}(v_2)\,. \tag{5.5}$$

Reality of the representation requires that $v_1 > -\alpha(2\gamma+1)$. Moreover, $\det[\Im(\varepsilon_N)] = 0$ relates $N$ values of the two parameters $v_1$ and $v_2$ (the parameter spectrum) for obtaining a solution at $\varepsilon = \varepsilon_N$. As examples, if we write $v_1 = -\alpha(2\gamma+1)+\alpha\xi$, then for $\varepsilon = \varepsilon_2 = (\gamma+1)^2$ a solution is obtained for $v_2^2 = \alpha\xi(\gamma+3/4)$, whereas for $\varepsilon = \varepsilon_3 = (\gamma+3/2)^2$ a solution is obtained for $v_2^2 = \alpha\left[2\xi(\gamma+1) + (\xi+1)(\gamma+5/4)\right]$.

In the Appendix we list, without giving details, few other examples of quasi-exactly solvable potentials indicating the class and subclass to which they belong. The square integrable basis and associated recursion relations are also given for each potential. One final remark is that these developments could be extended to relativistic quantum mechanics in a straight-forward manner. Currently, work is in progress to implement the same approach to the Dirac equation with various potential couplings including vector, scalar, and pseudo-scalar.

## ACKNOWLEDGMENTS

The author is grateful to M. E. H. Ismail for pointing out the problem and for stimulating and fruitful discussions. Suggestions made by R. S. Alassar for solutions of the integral equations in the Appendix are highly appreciated.

## APPENDIX
## FURTHER EXAMPLES

No general criteria were given for the selection of basis in the wavefunction expansion, beyond square integrability and compatibility with the boundary conditions, except for the requirements listed as Eqs. (2.4). Nonetheless, it is easier to give an ansatz for the weight function $\omega(y)$. Typically, for a system with bounded configuration space, $x \in [x_-, x_+]$, it could be written as $\omega(y) = y^{\gamma}(1-y)^{\alpha}(1+y)^{\beta}$ with some conditions on the real parameters *α*, *β*, and *γ*. This could always be done by rescaling to $x \in [-1, +1]$. On the other hand, for an infinite or semi-infinite coordinate space, the weight function could be written as $\omega(y) = y^{\gamma} e^{-\alpha y^{\beta}}$ with a proper choice of parameters. However, it is more difficult to propose a coordinate transformation function *y*(*x*) that would be compatible with the requirements in Eqs. (2.4). Nevertheless, we will attempt to give a particular formal solution as follows. Equation (2.4a) could be rewritten as

$$\frac{dy}{dx} = y\sqrt{A_0 y^{\sigma} + A_+ y^{\sigma+\mu} + A_- y^{\sigma-\mu}}\,. \tag{A.1}$$

Let's assume that we can write

1) For $\sigma = 0$: $A_0 + A_+ y^{+\mu} + A_- y^{-\mu} = \left(\tau_0 y^{\mu/2} + \eta_0 y^{-\mu/2}\right)^2$, (A.2a)

2) For $\sigma = +\mu$: $A_0 y^{+\mu} + A_+ y^{+2\mu} + A_- = \left(\tau_+ y^{\mu} + \eta_+\right)^2$, (A.2b)

3) For $\sigma = -\mu$: $A_0 y^{-\mu} + A_+ + A_- y^{-2\mu} = \left(\tau_- y^{-\mu} + \eta_-\right)^2$, (A.2c)

for some parameters $\{\tau_0, \tau_\pm, \eta_0, \eta_\pm\}$. Then *y*(*x*) could be written as solutions to the following nonlinear integral equations:

1) For $\sigma = 0$: $\tau_0 \int y^{1+\frac{\mu}{2}} dx + \eta_0 \int y^{1-\frac{\mu}{2}} dx = y$, (A.3a)

2) For $\sigma = +\mu$: $\tau_+ \int y^{1+\mu} dx + \eta_+ \int y\, dx = y$, (A.3b)

3) For $\sigma = -\mu$: $\tau_- \int y^{1-\mu} dx + \eta_- \int y\, dx = y$, (A.3c)

respectively. However, lacking the experience in the solution of such equations we will be contented with a trial-and-error scheme in proposing the following two sets of examples. Each set starts with a choice of basis followed by a list of all non-empty classes of quasi-exact solutions corresponding to $\sigma = 0, \pm\mu$. For each class we give the associated potential function and coefficients of the three-term recursion relation. Subsequently, the two solvability constraints are imposed and the specific recursion relation for that subclass is written down explicitly. Additionally, the orthogonality relation is given for each case that the polynomials form an orthogonal set. Reference is made in the examples to the following *N*×*N* tridiagonal symmetric matrix

$$J = \begin{pmatrix} j_0 & k_0 & & & & & \\ k_0 & j_1 & k_1 & & \mathbf{0} & & \\ & k_1 & j_2 & k_2 & & & \\ & & k_2 & \times & \times & & \\ & & & \times & \times & \times & \\ & \mathbf{0} & & & \times & j_{N-2} & k_{N-2} \\ & & & & & k_{N-2} & j_{N-1} \end{pmatrix} \tag{A.4}$$

**Examples (I):** $y(x)=1-2e^{-x}$, $\omega(y)=y^{\gamma}(1-y)^{\alpha}$: $x\in[0,\infty]\to y\in[-1,+1]$

**I.1)** $\sigma=-\mu=-1$, $\boxed{v(x)=\dfrac{v_1}{1-2e^{-x}}+\dfrac{v_2}{(1-2e^{-x})^2}}$:

$$C_0^{v}=-2\gamma\left(\gamma+\alpha-1/2\right)-v_1 \tag{A.5a}$$

$$\tilde{a}_n=2n\left(n+2\gamma+\alpha-1/2\right) \tag{A.5b}$$

$$d_n^{+}=-\left(n+\gamma+1/2\right)^2+v_2+\tfrac{1}{4} \tag{A.5c}$$

$$d_n^{-}=-\left(n+\gamma+\alpha-1\right)^2 \tag{A.5d}$$

I.1.1) $d_{N-1}^{+}=0\to v_2=(N+\gamma)(N+\gamma-1)$:

$$\begin{aligned}-v_1p_n(\varepsilon)&=2(n+\gamma)\left(n+\gamma+\alpha-\tfrac{1}{2}\right)p_n(\varepsilon)\\&+(N-n-1)(N+n+2\gamma)p_{n+1}(\varepsilon)+\left[\varepsilon-(n+\gamma+\alpha-1)^2\right]p_{n-1}(\varepsilon)\end{aligned} \tag{A.6}$$

Energy spectrum is solution of $\det\left[J(\varepsilon)\right]=0$, where

$$j_n=v_1+2(n+\gamma)\left(n+\gamma+\alpha-\tfrac{1}{2}\right) \tag{A.7a}$$

$$k_n^2=(N-n-1)(N+n+2\gamma)\left[\varepsilon-(n+\gamma+\alpha)^2\right] \tag{A.7b}$$

I.1.2) $\varepsilon=-d_N^{-}\to\varepsilon_N=-d_N^{-}=(N+\gamma+\alpha-1)^2$, $v_2>(N+\gamma)(N+\gamma-1)\equiv\tilde{v}$:

$$\begin{aligned}-v_1p_n(v_1)&=2(n+\gamma)\left(n+\gamma+\alpha-\tfrac{1}{2}\right)p_n(v_1)\\&+\left[(N-n-1)(N+n+2\gamma)+\xi^2\right]p_{n+1}(v_1)\\&+(N-n)(N+n+2\gamma+2\alpha-2)p_{n-1}(v_1)\quad,\end{aligned} \tag{A.8}$$

where $v_2=\tilde{v}+\xi^2$.

Parameter spectrum is $v_1=v_1(\xi,\gamma,\alpha,N)$ = – eigenvalues of $J$ with

$$j_n=2(n+\gamma)\left(n+\gamma+\alpha-\tfrac{1}{2}\right) \tag{A.9a}$$

$$k_n^2=(N-n-1)(N+n+2\gamma+2\alpha-1)\left[(N-n-1)(N+n+2\gamma)+\xi^2\right] \tag{A.9b}$$

$$\begin{aligned}\int\rho(v_1)p_n(v_1)p_m(v_1)dv_1&=(-)^n\tfrac{\Gamma(N-n)\Gamma(N+2\gamma+2\alpha-1)}{\Gamma(N)\Gamma(N+n+2\gamma+2\alpha-1)}\\&\times\tfrac{\Gamma(n+\gamma+\tau+1/2)\Gamma(n+\gamma-\tau+1/2)}{\Gamma(\gamma+\tau+1/2)\Gamma(\gamma-\tau+1/2)}\delta_{nm},\end{aligned} \tag{A.10}$$

where $\tau=\sqrt{\left(N+\gamma-1/2\right)^2+\xi^2}$ and $n,m\le N-1$.

**Examples (II):** $y(x)=\text{sech}(x)$, $\omega(y)=y^{\gamma}\left(1-y^2\right)^{\alpha/2}$: $x\in[0,\infty]\to y\in[0,+1]$

**II.1)** $\sigma=0$, $\mu=2$, $\boxed{v(x)=\alpha(\alpha-1)\,\text{csch}(x)^2+v_1\,\text{sech}(x)^2+v_2\cosh(x)^2}$:

$$a_n=(2n+\gamma)^2 \tag{A.11a}$$

$$d_n^{+}=v_2 \tag{A.11b}$$

$$d_n^{-}=(2n+\gamma+\alpha-1)(2n+\gamma+\alpha-2)+v_1 \tag{A.11c}$$

II.1.1) $d_{N-1}^{+}=0\ \rightarrow\ v_2=0$: Exactly solvable Pöschl-Teller.

II.1.2) $d_N^{-}=0\ \rightarrow\ v_1=-(2N+\gamma+\alpha-1)(2N+\gamma+\alpha-2)$

$$\varepsilon p_n(\varepsilon)=-(2n+\gamma)^2 p_n(\varepsilon)+v_2\, p_{n+1}(\varepsilon)$$
$$-4(N-n)\left(N+n+\gamma+\alpha-\tfrac{3}{2}\right)p_{n-1}(\varepsilon) \tag{A.12}$$

where $v_2<0$.

$$\int\rho(\varepsilon)p_n(\varepsilon)p_m(\varepsilon)d\varepsilon=\left(-4/v_2\right)^n\frac{\Gamma(N-n)\Gamma(N+\gamma+\alpha-1/2)}{\Gamma(N)\Gamma(N+n+\gamma+\alpha-1/2)}\delta_{nm}\,,\ \ n,m\le N-1 \tag{A.13}$$

Energy spectrum is the set of eigenvalues of $J$ with

$$j_n=-(2n+\gamma)^2 \tag{A.14a}$$
$$k_n^2=-4v_2(N-n-1)\left(N+n+\gamma+\alpha-\tfrac{1}{2}\right) \tag{A.14b}$$

**II.2)** $\sigma=\mu=1$, $\boxed{v(x)=\alpha(\alpha-1)\,\mathrm{csch}(x)^2+v_1\,\mathrm{sech}(x)^2+v_2\,\mathrm{sech}(x)}$:

$$C_0^v=-v_2 \tag{A.15a}$$
$$\tilde{a}_n=0 \tag{A.15b}$$
$$d_n^{+}=-(n+\gamma+1)^2 \tag{A.15c}$$
$$d_n^{-}=(n+\gamma+\alpha)\left(n+\gamma+\alpha-1\right)+v_1 \tag{A.15d}$$

II.2.1) $d_N^{-}=0\ \rightarrow\ v_1=-(N+\gamma+\alpha)(N+\gamma+\alpha-1)$:

$$v_2\, p_n(\varepsilon)=(N-n)(N+n+2\gamma+2\alpha-1)p_{n-1}(\varepsilon)+\left[(n+\gamma+1)^2-\varepsilon\right]p_{n+1}(\varepsilon)\,, \tag{A.16}$$

where $\varepsilon\le\hat{\varepsilon}=(\gamma+1)^2$.

Energy spectrum is solution of $\det\left[J(\varepsilon)\right]=0$, where

$$j_n=-v_2 \tag{A.17a}$$
$$k_n^2=(N-n-1)(N+n+2\gamma+2\alpha)\left[(n+\gamma+1)^2-\varepsilon\right] \tag{A.17b}$$

II.2.2) $\varepsilon=-d_{N-1}^{+}\ \rightarrow\ \varepsilon_N=-d_{N-1}^{+}=(N+\gamma)^2$, $v_1>-(\gamma+\alpha)(\gamma+\alpha+1)\equiv\tilde{v}$:

$$-v_2 p_n(v_2)=(N-n-1)(N+n+2\gamma+1)p_{n+1}(v_2)$$
$$+\left[(n-1)(n+2\gamma+2\alpha)+\xi^2\right]p_{n-1}(v_2)\,, \tag{A.18}$$

where $v_1=\tilde{v}+\xi^2$.

Parameter spectrum is $v_1=v_1(\xi,\gamma,\alpha,N)$ = – eigenvalues of $J$ with

$$j_n=0 \tag{A.19a}$$
$$k_n^2=(N-n-1)(N+n+2\gamma+1)\left[n(n+2\gamma+2\alpha+1)+\xi^2\right] \tag{A.19b}$$

$$\int\rho(v_2)p_n(v_2)p_m(v_2)dv_2=\frac{\Gamma(N-n)\Gamma(N+2\gamma+1)}{\Gamma(N)\Gamma(N+n+2\gamma+1)}$$
$$\times\frac{\Gamma(n+\gamma+\alpha+\tau+1/2)\Gamma(n+\gamma+\alpha-\tau+1/2)}{\Gamma(\gamma+\alpha+\tau+1/2)\Gamma(\gamma+\alpha-\tau+1/2)}\delta_{nm}\,, \tag{A.20}$$

where $\tau=\sqrt{\left(\gamma+\alpha+1/2\right)^2-\xi^2}$ and $n,m\le N-1$.

**II.3)** $\sigma=\mu=2$, $\boxed{v(x)=\alpha(\alpha-1)\,\mathrm{csch}(x)^2+v_1\,\mathrm{sech}(x)^2+v_2\,\mathrm{sech}(x)^4}$:

$$C_0^v=-(\gamma+\alpha)\left(\gamma+\alpha+1\right)-v_1 \tag{A.21a}$$
$$\tilde{a}_n=4n\left(n+\gamma+\alpha+1/2\right) \tag{A.21b}$$
$$d_n^{+}=-\left(2n+\gamma+2\right)^2 \tag{A.21c}$$

$$d_n^- = v_2 \tag{A.21d}$$

II.3.1) $d_N^- = 0 \rightarrow v_2 = 0$: Exactly solvable Pöschl-Teller.

II.3.2) $\varepsilon = -d_{N-1}^+ \rightarrow \varepsilon_N = -d_{N-1}^+ = (2N+\gamma)^2$, $v_2 > 0$:

$$-v_1 p_n(v_1) = (2n+\gamma+\alpha)(2n+\gamma+\alpha+1)p_n(v_1) + v_2\, p_{n-1}(v_1)$$
$$+4(N-n-1)(N+n+\gamma+1)p_{n+1}(v_1) \tag{A.22}$$

Parameter spectrum is $v_1 = v_1(v_2,\gamma,\alpha,N)$ = – eigenvalues of *J* with

$$j_n = (2n+\gamma+\alpha)(2n+\gamma+\alpha+1) \tag{A.23a}$$

$$k_n^2 = 4v_2(N-n-1)(N+n+\gamma+1) \tag{A.23b}$$

$$\int \rho(v_1)\, p_n(v_1)\, p_m(v_1)\, dv_1 = \left(4/v_2\right)^n \frac{\Gamma(N)\Gamma(N+n+\gamma+1)}{\Gamma(N-n)\Gamma(N+\gamma+1)}\delta_{nm} \tag{A.24}$$

**FIGURES CAPTION:**

**Fig. 1:** Graph of the weight function $\rho(\varepsilon)$ associated with the orthogonality relation (3.13) for $N$ = 10, $\alpha=\frac{1}{4}$, $\gamma=1$, and $v_2=\frac{1}{2}$ (i.e., $\xi=11/8$).

**Fig. 2** (color online): Plot of few low degree polynomials satisfying the recursion relation (3.9b) with the same physical parameters as those in Fig. 1 but for $N$ = 7. The energy eigenvalues $\{\varepsilon_n\}_{n=0}^{6}$ are shown as solid black circles on the energy axis.

**Fig. 3:** The weight function $\rho(\varepsilon)$ associated with the orthogonality relation (3.20) for $N$ = 20, $\alpha=\xi=20$, and $\gamma=1$.

**Fig. 4** (color online): Few of the low degree polynomials satisfying the recursion relation (3.18)' for the same physical parameters as those in Fig. 3 but for $N$ = 7. The energy eigenvalues $\{\varepsilon_n\}_{n=0}^{6}$ are shown as solid black circles on the energy axis.

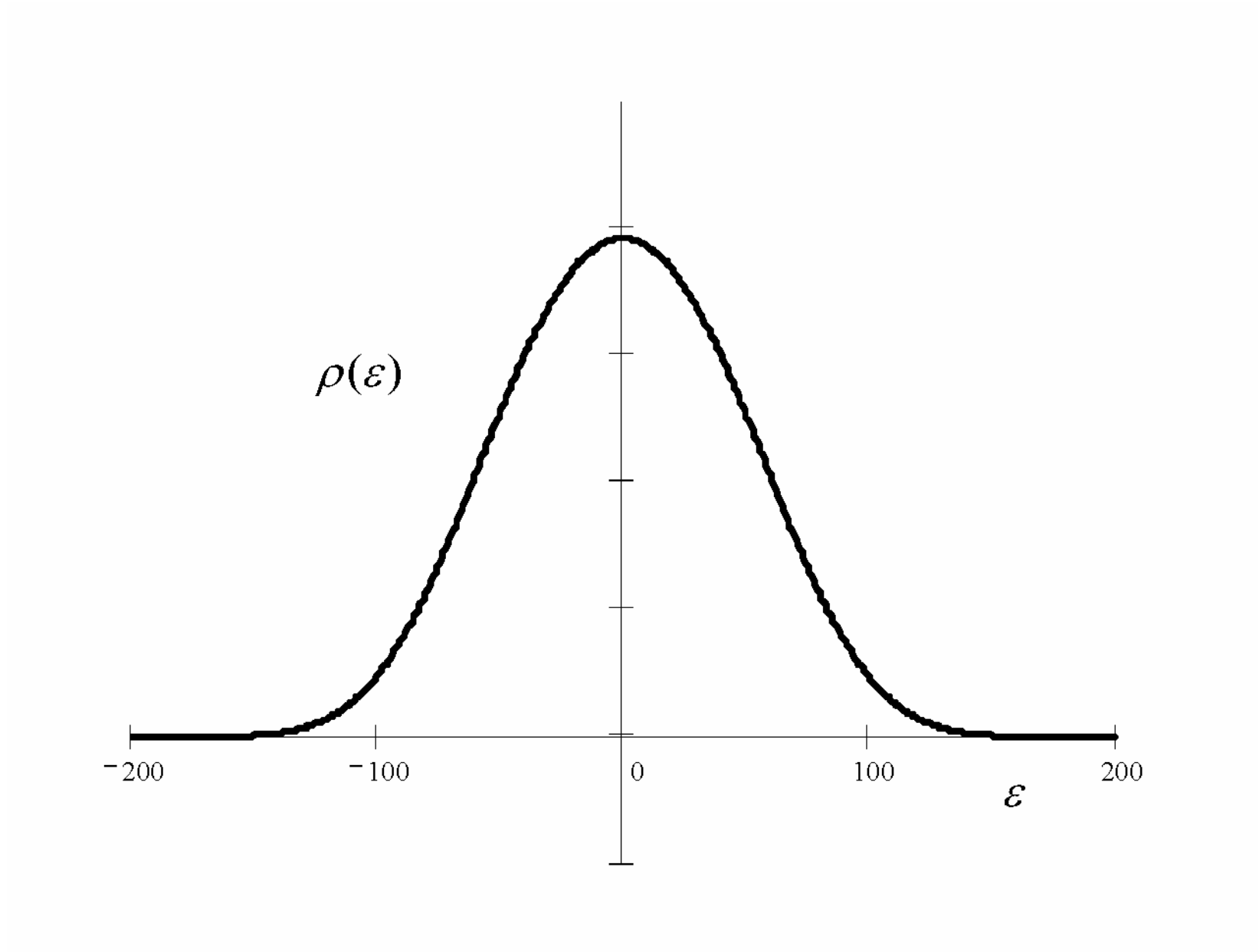

Fig. 1

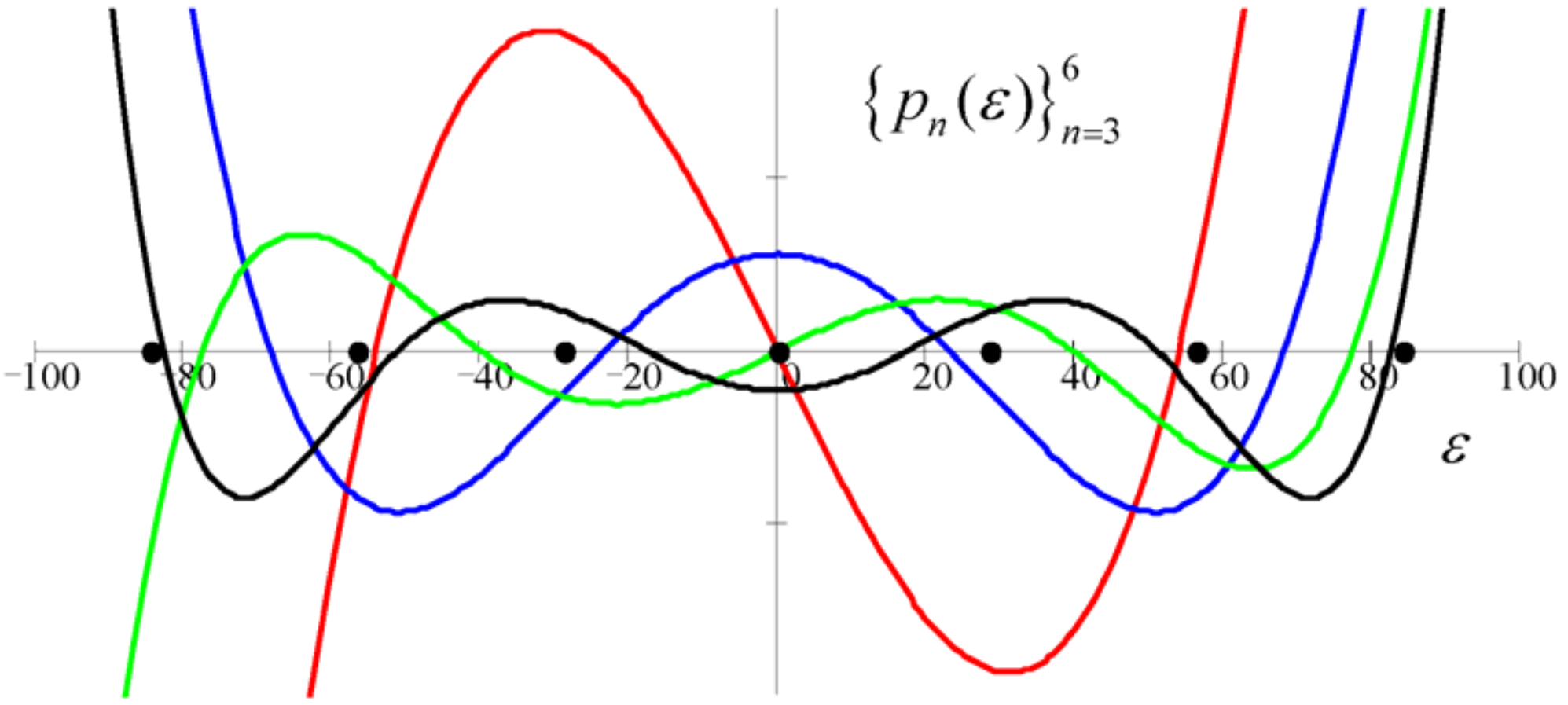

Fig. 2

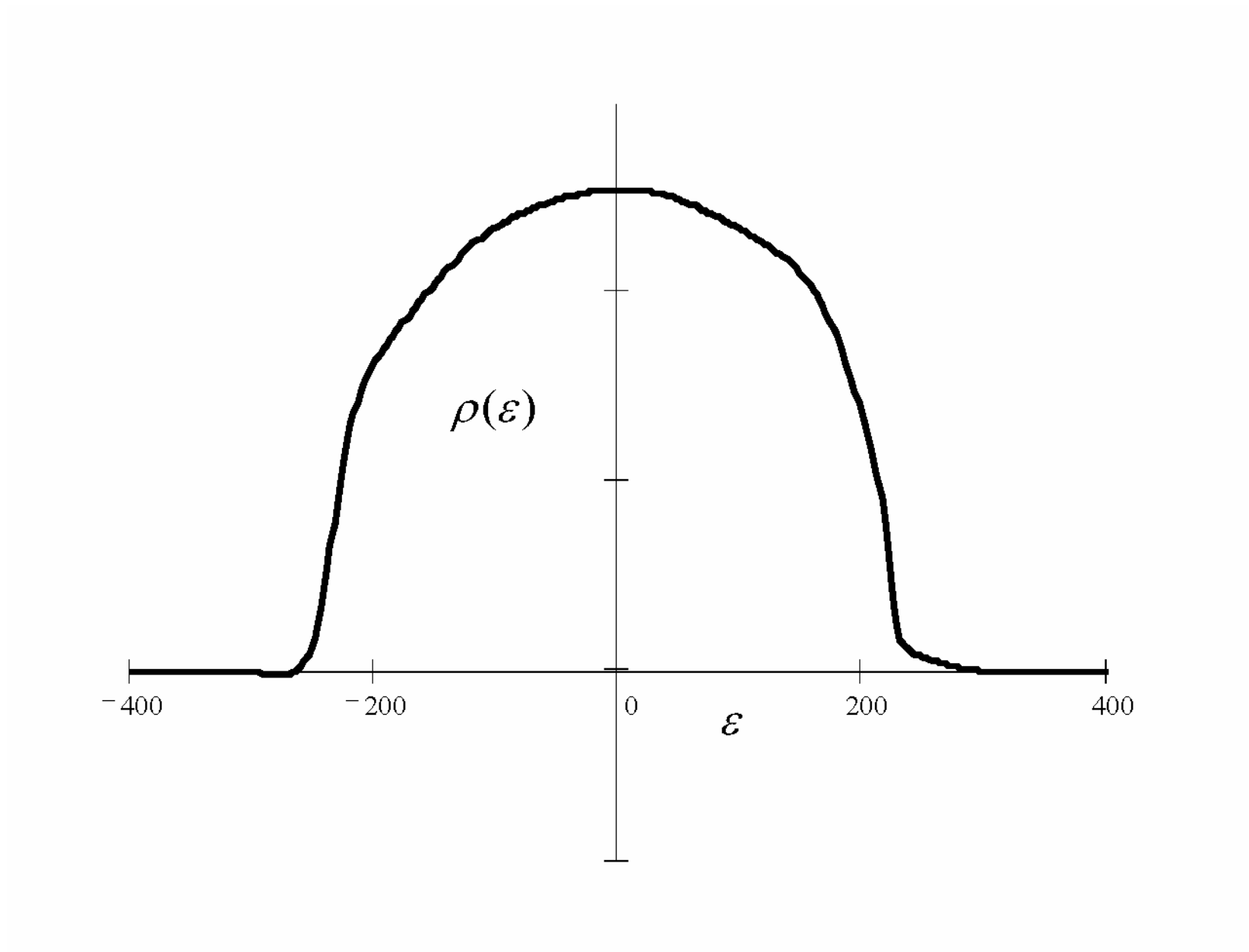

Fig. 3

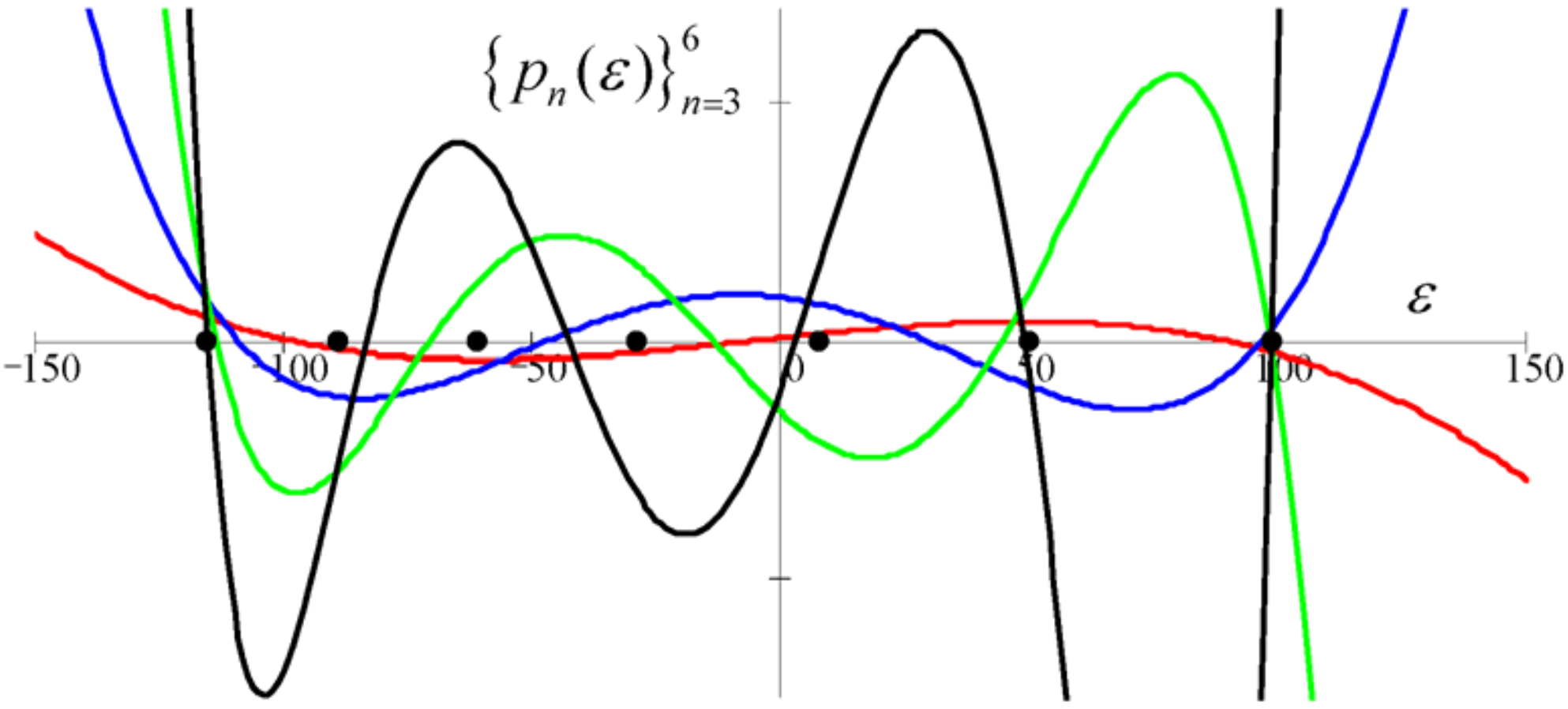

$\{p_n(\varepsilon)\}_{n=3}^{6}$
$\varepsilon$
−150
150

Fig. 4